\documentclass[]{aa}
\usepackage{txfonts}
\usepackage{natbib}
\usepackage{graphicx}
\usepackage{longtable}
\bibpunct{(}{)}{;}{a}{}{,}
\usepackage{color}

\begin{document}

\title{Investigating potential planetary nebula/cluster pairs\thanks{Based on observations gathered with ESO-VISTA
telescope (program ID 172.B-2002)}
\thanks{Based on observations gathered at Las Campanas observatory (program ID CN2012A-080).}
\thanks{The spectra are available at the CDS via anonymous ftp to cdsarc.u-strasbg.fr (130.79.128.5) or via
http://cdsweb.u-strasbg.fr/cgi-bin/qcat?J/A+A/.}
}
\subtitle{}

\author{
C. Moni Bidin\inst{1,2}
\and
D. Majaess\inst{3}
\and
C. Bonatto\inst{4}
\and
F. Mauro\inst{1}
\and
D. Turner\inst{5}
\and
D. Geisler\inst{1}
\and
A.-N. Chen\'e\inst{1,6}
\and
A. C. Gormaz-Matamala\inst{1}
\and
J. Borissova\inst{6}
\and
R. G. Kurtev\inst{6}
\and
D. Minniti\inst{7,8}
\and
G. Carraro\inst{9,10}
\and
W. Gieren\inst{1}
}

\institute{
Departamento de Astronom\'ia, Universidad de Concepci\'on, Casilla 160-C, Concepci\'on, Chile
\and
Instituto de Astronom\'ia, Universidad Cat\'olica del Norte, Av. Angamos 0610, Antofagasta, Chile
\and
Halifax, Nova Scotia, Canada B3K 5L3
\and
Departamento de Astronomia, Universidade Federal do Rio Grande do Sul, Av. Bento Gon\c{c}alves 9500 Porto Alegre
91501-970, RS, Brazil
\and
Department of Astronomy and Physics, Saint Mary's University, Halifax, NS B3H 3C3, Canada
\and
Departamento de F\'isica y Astronom\'ia, Facultad de Ciencias, Universidad de Valpara\'iso, Av.
Gran Breta\~na 1111, Valpara\'iso, Chile
\and
Departamento de Astronom\'ia y Astrof\'isica, Pontificia Universidad Cat\'olica de Chile, Casilla 306,
Santiago, Chile
\and
Vatican Observatory, V 00120, Vatican City State, Italy
\and
ESO - European Southern Observatory, Avda Alonso de Cordova, 3107, Casilla 19001, Santiago, Chile
\and
Dipartimento di Fisica e Astronomia, Universit\'a di Padova, Via Marzolo 8, I-35131 Padova, Italy
}
\date{Received / Accepted }


\abstract
{Fundamental parameters characterizing the end-state of intermediate-mass stars may be constrained by discovering planetary
nebulae (PNe) in open clusters (OCs). Cluster membership may be exploited to establish the distance, luminosity, age, and
physical size for PNe, and the intrinsic luminosity and mass of its central star.}
{Four potential PN-OC associations were investigated to assess the cluster membership for the PNe.}
{Radial velocities were measured from intermediate-resolution optical spectra, complemented with previous estimates in the
literature. When the radial velocity study supported the PN/OC association, we analyzed whether other parameters (e.g., age,
distance, reddening, central star brightness) were consistent with this conclusion.}
{Our measurements imply that the PNe VBe\,3 and HeFa\,1 are not members of the OCs NGC\,5999 and NGC\,6067, respectively,
and that they very likely belong to the background bulge population. Conversely, consistent radial velocities indicate that
NGC\,2452/NGC\,2453 could be associated, but our results are not conclusive so additional observations are warranted.
Finally, we demonstrate that all the available information point to He\,2-86 being a young, highly internally obscured PN
member of NGC\,4463. New near-infrared photometry acquired via the Vista Variables in the Via Lactea ESO public survey was
used in tandem with existing $UBV$ photometry to measure the distance, reddening, and age of NGC\,4463, finding
$d=1.55\pm0.10$~kpc, $E(B-V)=0.41\pm0.02$, and $\tau=65\pm10$~Myr, respectively. The same values should be adopted for
the PN if the proposed cluster membership is confirmed.}
{}


\authorrunning{Moni Bidin et al.}
\mail{cmoni@ucn.cl}
\titlerunning{}
\maketitle


\section{Introduction}
\label{s_intro}

Our knowledge of the intrinsic properties of the Galactic planetary nebulae (PNe) has been restricted in part by large
uncertainties tied to their distance estimates, accurate only within 35\% to 50\% on average \citep{Zhang95}. That
unsatisfactory uncertainty is confirmed by Fig. 6 in \citet{Giammanco11}, find significant (random and
systematic) offsets between the latest IPHAS extinction distances for PNe and existing estimates. The impetus for
discovering PNe in open clusters (OCs) is thus clear, because well-studied OCs possessing solid spectroscopic
observations and deep multi-band photometry may yield distance uncertainties less than 10\%
\citep[e.g.,][]{Majaess12,Chene12}. PNe in OCs will always be a rare phenomenon representing only a tiny
fraction of the $\sim$3000 known Galactic PNe, but they may be subsequently employed as crucial calibrators for
methods used to establish PNe distances \citep{Bensby01,Osterbrock06,Frew06}.

Establishing cluster membership for PNe offers a potential means of determining their fundamental properties. A PN's true
dimensions and age can be deduced from cluster membership, given the availability of an apparent angular diameter and
expansion velocity for the object. Moreover, main-sequence progenitor masses for PNe may be constrained by examining
cluster members near the turnoff. Unfortunately, detection of an association between a PN and an OC is hampered by
numerous factors, including the short lifetimes of PNe. For example, for a Galactic OC exhibiting an age of 100~Myr,
the main-sequence mass for the PN progenitor is $\sim$4~M$_\odot$, which implies a PN phase that lasts for merely
$\sim10^{3}$~years. In general, that lifetime is sensitive to the progenitor mass and mass loss along the red giant
branch \citep[see, e.g.,][]{Schoenberner96}, and typically ranges from 10$^{3}$ to 10$^{5}$ years
\citep{Schoenberner96,Koppen00}. \citet{Jacob13} have recently proposed that the typical visibility time of PN is only
$\sim 20\,000$~years. Furthermore, the rapid dissolution rate of OCs ensures that a significant fraction of OCs
capable of housing a PN have dissolved by the time they are old enough to produce one
\citep{Battinelli91,Lada03,delaFuente08}. Younger OCs exhibiting ages less than $\sim$30~Myr are probably excluded
because their evolved stars are believed to terminate as Type-II supernovae. Moreover, \citet{Larsen06} detected
only three PNe in their sample of eighty extragalactic OCs, an occurrence a factor of two lower than expected. They
suggest that the discrepancy could be due, among other factors, to uncertainties in PN lifetimes.

\citet{Majaess07} discuss the possible cluster membership of thirteen planetary nebulae. Chance coincidences were
found to exist for seven of the cases considered, but the authors advocated follow-up studies for at least six PN/OC pairs
in which a physical association was not excluded by the available evidence, namely M\,1-80/Berkeley\,57,
NGC\,2438/NGC\,2437, NGC\,2452/NGC\,2453, VBRC\,2 and NGC\,2899/IC\,2488, and HeFa\,1/NGC\,6067. \citet{Majaess07}
likewise tabulated a number of additional potential associations between planetary nebulae and open clusters, which
included G305.3$-$03.1/Andrews-Lindsay\,1 (ESO\,96$-$SC04) and He\,2-86/NGC\,4463. 

Numerous studies have since been published that
have advanced the field and helped clarify the situation concerning several potential PN/OC associations. A follow-up study by
\citet{Kiss08} contradicted the previous results of \citet{Pauls96}, concluding that AAOmega radial velocities (RVs) rule
out membership for NGC\,2438 in NGC\,2437, since exists an offset of $\sim30$~km~s$^{-1}$ exists between the objects.
\citet{Bonatto08} reports the discovery of a new open cluster (designated Bica\,6) and proposed that it hosts the PN
Abell\,8, on the basis of parameters inferred from Two Micron All Sky Survey \citep[2MASS,][]{Skrutskie06} near-infrared
colour-magnitude diagrams (CMDs), and stellar radial density profiles. \citet{Turner11} subsequently obtained CCD spectra,
$UBVRI_C$ photometry, and RVs for luminous cluster stars in Bica\,6 to complement the \citet{Bonatto08} analysis, and
reaffirmed that the PN is a cluster member. \citet{Parker11} secured new RVs for the PN
G305.3$-$03.1 \citep[PHR 1315$-$6555,][]{Parker06} and members of Andrews-Lindsay\,1.

By comparing the cluster and
PN parameters, they demonstrate that all available evidence points to a physical association between the objects. However,
the distance estimates for that heavily reddened faint cluster vary from 7.57~kpc \citep{Phelps94} to 16.9~kpc
\citep{Carraro05}. The sizable dust extinction along the line of sight to Andrews-Lindsay\,1 was most likely a contributing
factor that hindered the establishment of a precise cluster distance, especially since existing results are mainly tied
to optical observations. Additional observations are desirable for determining the cluster distance with better accuracy so as
to eventually take advantage of the proven cluster membership to study the PN properties in detail.

A close apparent spatial coincidence is clearly only a first step toward identifying physically related PN/OC pairs. A common
RV is also required, as an additional, necessary but not sufficient condition. Ideally, the independently measured distance
and reddening of both the PN and OC, in addition to RVs, should show a good match to prove their association beyond doubt.
Unfortunately, as already discussed, the distance estimates of most PNe are based on statistical methods that are not
particularly reliable for a single object. As a consequence, the literature values are often uncertain, and very discrepant
(see, for example, the case of He\,2-86 discussed in Table~\ref{t_dPN}). On the other hand, the reddening of a PN must
be used with caution, since PNe could suffer from non-negligible internal dust extinction. This is still a controversial point
in the literature. For example, \citet{Gathier86} claim that internal reddening of PNe rarely exceeds 0.05~magnitudes, while
\citet{Zagury05} suggests that 0.1--0.2~magnitudes could be common, and much higher values can be found in a few cases.
\citet{Phillips98} also argues that internal dust opacity in younger nebulae is often appreciable. Thus, a comparison of all
the PN and OC parameters can be far from straightforward.

This study constitutes the first in an extensive project undertaken to assess membership for PNe lying in close apparent
angular proximity to open clusters. Four potential PN/OC pairs are investigated in this study, namely
VBe\,3(G326.1$-$01.9)/NGC\,5999, HeFa\,1(G329.5$-$02.2)/NGC\,6067, NGC\,2452/NGC\,2453, and
He\,2-86(G300.7$-$02.0)/NGC\,4463. Aware that many proposed associations are actually chance alignments, we
preliminarily measured RVs on intermediate-resolution spectra, to identify the pairs worthy of more time-consuming, follow-up
investigations.


\section{Spectroscopic data and measurements}
\label{s_spectra}

\begin{figure}
\begin{center}
\includegraphics[width=9.5cm]{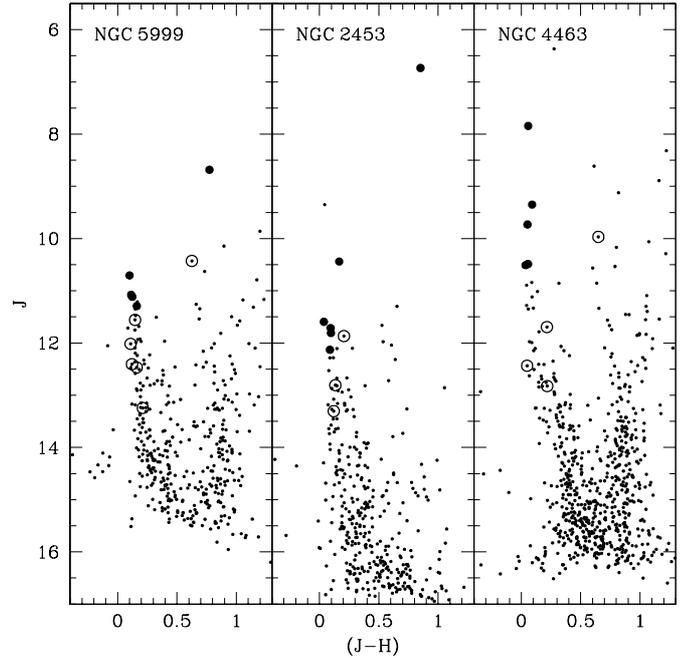}
\caption{Color-magnitude diagrams of the program clusters. The 2MASS point sources within $3\arcmin$ from the cluster
center are plotted as small dots. Full circles indicate the selected targets, while empty circles convey the positions
of stars that entered the slit fortuitously.}
\label{f_tarsel}
\end{center}
\end{figure}

Our observations targeted three OCs and three PNe of the four PN/OC pairs object of the present work. Precise RV of the
PN NGC\,2452 and the OC NGC\,6067 have previously been measured by other authors, so we took these values from the
literature.

The program stars were selected from the 2MASS catalog. The criteria for selection were the proximity to the cluster
evolutionary sequence in the IR color-magnitude diagram (CMD), the stellar magnitude, and the distance from the cluster
center. The positions of the observed stars in the CMD are indicated in Fig.~\ref{f_tarsel}, and their photometric data
are provided in Table~\ref{t_RVs}.

The spectra were collected on 2012 April 13 with the B\&C spectrograph at the focus of the 2.5m du~Pont telescope at
the Las Campanas observatory. The 1200~line/mm grating was used at the second order, with a grating angle of $56\fdg 12$,
thus collecting data in the spectral interval 5960--6660~\AA. This range was selected to target various features
suitable for inferring radial velocity measurements in the spectra of hot stars, red giants, and PNe. In fact, the
spectral range could not be changed during observations owing to the non-repeatability of the exact grating angle position.
The slit width was adjusted to the instrumental anamorphic demagnification \citep{Schweizer79} to have a 2.5-pixel
resolution of 0.85~\AA\ on the CCD. The corresponding projected width on the sky was $1\farcs3$. All the selected
targets, both stars and nebulae, were centered on the slit and in the same position along the spatial direction, to avoid
systematics between the measurements. This point is particularly relevant for PNe, whose off-center RV can differ from the
systemic value. During daytime operations prior to the run, we realized that focusing the instrument throughout the entire
spectral range was impossible. We therefore decided to focus the redder half of the spectra, which hosts the H$_\alpha$
line, since this feature is suitable for inferring radial velocity measurements for the PNe and program stars. As a
consequence, the spectra was progressively de-focused toward shorter wavelengths, but the present investigation was
entirely restricted to the well-focused part redward of 6300~\AA.

\begin{table}[t]
\begin{center}
\caption{Log of the observations. ``wlc'' refers to the arc exposures.}
\label{t_obslog}
\begin{tabular}{c c c c}
\hline
\hline
target & cluster & texp (s) & UT (start) \\
\hline
TYC\,6548-790-1 & NGC\,2453 & \ \ 600 & 23:48:48 \\
MSP\,159 & NGC\,2453 & \ \ 900 & 00:04:39 \\
wlc1 & & & 00:20:30 \\
TYC\,6548-905-1 & NGC\,2453 & \ \ 900 & 00:27:06 \\
TYC\,6548-507-1 & NGC\,2453 & \ \ 900 & 00:44:52 \\
TYC\,6548-1643-1 & NGC\,2453 & \ \ 900 & 01:02:52 \\
MSP\,111 & NGC\,2453 & \ \ 900 & 01:27:55 \\
wlc2 & & & 01:43:47 \\
HD\,108719 & NGC\,4463 & \ \ 300 & 01:52:30 \\
CPD$-$64~1946 & NGC\,4463 & \ \ 600 & 02:00:57 \\
2MASS\,12294384$-$6448130 & NGC\,4463 & \ \ 600 & 02:14:09 \\
wlc3 & & & 02:25:22 \\
CPD$-$64 1944 & NGC\,4463 & \ \ 900 & 02:35:06 \\
wlc4 & & & 02:55:01 \\
He\,2-86 & PN & \ \ 600 & 05:15:57 \\
wlc5 & & & 05:27:09 \\
2MASS\,12294705$-$6447464 & NGC\,4463 & \ \ 900 & 05:32:54 \\
VBe\,3 & PN & 1200 & 06:19:07 \\
wlc6 & & & 06:40:44 \\
HeFa\,1 & PN & 1800 & 07:08:58 \\
wlc7 & & & 08:28:54 \\
2MASS\,J15521389$-$5628139 & NGC\,5999 & \ \ 750 & 08:36:37 \\
TYC\,8705-3241$-$1 & NGC\,5999 & \ \ 900 & 08:51:36 \\
wlc8 & & & 09:07:35 \\
2MASS\,J15521176$-$5628045 & NGC\,5999 & \ \ 900 & 09:11:14 \\
2MASS\,J15521358$-$5629086 & NGC\,5999 & \ \ 900 & 09:28:08 \\
2MASS\,J15522168$-$5626451 & NGC\,5999 & \ \ 900 & 09:47:47 \\
wlc9 & & & 10:03:17 \\
\hline
\end{tabular}
\end{center}
\end{table}

The log of the observations is presented in Table~\ref{t_obslog}. Exposure times varied between 300s and 900s,
depending on the stellar magnitude to ensure similar signal-to-noise (S/N) quality for all targets. The exposures for
the PNe were a factor of two longer and were adjusted to match the brightness of the H$_\alpha$ emission. Observing
numerous stars in the same cluster only required minimal telescope movement, and the instrumental conditions were expected
to remain comparatively stable. Consequently, a lamp frame for wavelength calibration (indicated as ``wlc" in
Table~\ref{t_obslog}) was collected each time the telescope was moved to observe a different cluster or if a given
cluster was observed longer than an hour. During data reduction, we verified that the shift between two consecutive lamp
exposures was typically 1--2~km~s$^{-1}$, except between wlc8 and wlc9, where it reached 4~km~s$^{-1}$. However, the
three stars observed between these lamp images do not show, within the errors, a systematic offset with respect to the other
stars observed in the same cluster (NGC\,5999).

\begin{figure}
\begin{center}
\includegraphics[width=9cm]{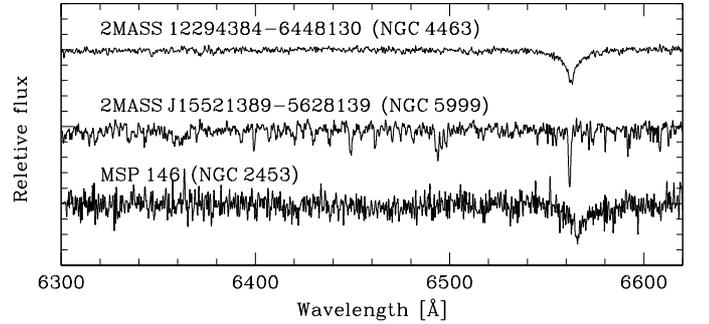}
\caption{Examples of the collected spectra. The spectra have been shifted vertically to avoid overlap.}
\label{f_spectraC}
\end{center}
\end{figure}

\begin{figure}
\begin{center}
\includegraphics[width=9cm]{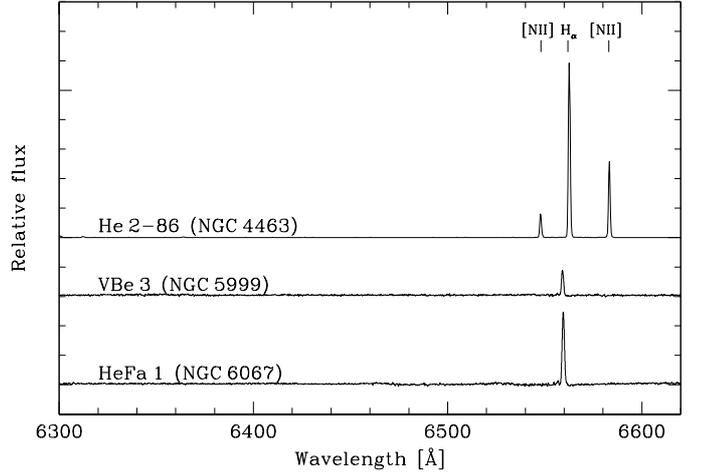}
\caption{Spectra of the target PNe. The spectra have been shifted vertically to avoid overlap, and that of
He\,2-86 was vertically reduced by a factor of ten.}
\label{f_spectraPN}
\end{center}
\end{figure}

The data were reduced by means of standard IRAF\footnote{IRAF is distributed by the National Optical Astronomy
Observatories, which are operated by the Association of Universities for Research in Astronomy, Inc., under
cooperative agreement with the National Science Foundation.} routines. The frames were debiased and flat-fielded using
calibration frames acquired during daytime operations, and the one-dimensional spectra were subsequently extracted with
an optimum algorithm \citep{Horne86}. The sky background was estimated from two adjacent regions on both sides of the
stellar spectrum and subtracted. For each star, the dispersion solution was obtained with the respective lamp frame,
extracted at the same position on the CCD as the science spectrum to avoid biases introduced by the curvature of the
arcs. The lamp emission lines in the out-of-focus bluer half of the spectrum were double-peaked. To overcome this
problem, the lamp spectrum was convolved with a Gaussian with FWHM=10 pixels. That procedure produced well-shaped
lines without altering the line centers and did not cause blending among the few well-separated features. The
dispersion solution was derived with a third-order polynomial fit, and the rms of the residuals was typically
$\sim$0.023~\AA, i.e. $\sim$1.1~km~s$^{-1}$ at H$_\alpha$. The lamp spectra were calibrated, and the position of the
lamp lines was compared to laboratory wavelengths to check the final calibration solution. Finally, the spectra were
normalized as required for the radial velocity measurements. Some extracted spectra are illustrated in
Fig.~\ref{f_spectraC} as an example, where we show a typical spectrum for a hot target (2MASS\,12294384$-$6448130), a
cool red giant (2MASS\,J15521389$-$5628139), and the noisy spectrum of a star that fell in the spectrograph slit by
chance (MSP\,146). The resulting S/N for the final spectra of the selected targets was typically in the range of
20--50. The spectra of the three program PNe are displayed in Fig.~\ref{f_spectraPN}.

Non-program stars fell in the spectrograph slit in each exposure, as expected by the nature of the surveyed
low-latitude crowded stellar fields near open clusters. Their spectra were extracted and analyzed in a similar fashion to
the main targets. Most of these objects were too faint to supply pertinent information, since the resulting
measurements were highly uncertain for them. The radial velocities established were different from that of the cluster
stars in most cases, indicating that they are probably field stars. However, the off-center position in the slit for such
objects may have introduced systematic errors in the radial velocity. The typical seeing during the run was
1$\arcsec$, slightly lower than the projected slit width, a fact that could have enhanced this well-known problem of
slit spectroscopy \citep[see, for example, the analysis in][]{Moni06}. In certain cases, their position in the CMD
merely confirmed their status as field stars. Still, the photometry and RV of some of these additional stars were
compatible with cluster membership, and the non-members helped distinguish the cluster RV from the field in one case
(Sect.~\ref{ss_4463}). They are therefore included in our analysis and are indicated with ``A" in the second column
of Table~\ref{t_RVs}, whereas the selected targets are marked with ``T".

RVs for the program stars were measured via cross-correlation \citep[CC,][]{Simkin74,Tonry79}, as
implemented in the {\it fxcor} IRAF task.  The peak of the CC function was fit with a Gaussian profile. Synthetic
spectra of solar metallicity drawn from the library of \citet{Coelho05} were used as templates. Previous investigations
have shown that the RV measurements are not affected by the exact choice of the template, because a partial mismatch
between object and template spectral types merely enhances the formal uncertainties, without shifting the peak of the CC
function \citep{Morse91,Moni11}. The spectra of hot stars, which show a broad and isolated H$_\alpha$ feature, were
cross-correlated with a spectrum with T$_\mathrm{eff}$=5500~K and $\log{g}$=4.0. A cooler model with
T$_\mathrm{eff}$=4500~K and $\log{g}$=2.0 was adopted for the few giant stars. The latter were recognized by the
narrowness of H$_\alpha$ and the numerous blended lines. In both cases, the CC was restricted to the range
6520--6590~\AA, comprising the H$_\alpha$ line and wings. The CC error was in the range of 3--6~km~s$^{-1}$ for the
well-exposed targets, while it increased to upward of 20--30~km~s$^{-1}$ for the fainter stars that happened to fall
within the slit. As a consistency check, we measured the RV of each target again by fitting the core of the H$_\alpha$
line with a Gaussian profile. The accuracy of these measurements was lower, typically by a factor of two, but the mean
difference with the CC results was $\overline{\Delta \mathrm{RV}}=-0.7$~km~s$^{-1}$ with an rms of 4.3~km~s$^{-1}$,
showing no systematic difference between our values and the results of this independent direct estimate.

RVs for the PNe were measured by cross-correlating their spectrum with a synthetic H$_\alpha$ emission profile,
convolved with a Gaussian to match the instrumental resolution of the observed spectra. The H$_\alpha$ emission line was
always bright and well exposed. We checked that our results are consistent within 1$\sigma$ with those obtained by fitting
the observed lines with a Gaussian profile, although this procedure returned final errors higher by a factor of two.
Additional emission lines at bluer wavelengths were visible in some cases, but they were not employed to infer the
radial velocity owing to their extreme faintness and the focus problem mentioned before. On the well-focused redder half,
bright [\ion{N}{ii}] lines at 6548 and 6583~\AA\ were observed in the spectrum of He\,2-86 in NGC\,4463. No signature
of the [WC4] central star was identified in the spectrum, whose features are likely to be too faint \citep{Acker03} for our
data. In this case, the RV was determined from the three observed features, cross-correlating the spectrum
in the range 6540--6590~\AA\ with a synthetic template comprising both the H$_\alpha$ and the two [\ion{N}{ii}] lines.

All spectra were extracted again without background subtraction. The position of the [\ion{O}{i}] sky emission lines at
6300.0 and 6363.8~\AA, measured by means of a Gaussian fit, was used to evaluate possible zero-point offsets. The
resulting corrections to the measured RVs were typically on the order of $-$10~km~s$^{-1}$. That procedure was performed
slightly off-center of the object spectrum in the case of PNe to avoid contamination by their emission lines. Finally,
the RVs were transformed to heliocentric velocities. The results are summarized in Table~\ref{t_RVs}. The errors were
obtained from the quadratic sum of the relevant uncertainties affecting
our procedure: the CC error, the wavelength calibration error, and the uncertainty in the zero-point correction.  This
last quantity typically amounted to 1--3~km~s$^{-1}$. The typical total error for target stars varied from 4 to
10~km~s$^{-1}$, except for one case.

\begin{table*}[t]
\begin{center}
\caption{Photometric data and radial velocities of the program objects. The cluster velocity adopted is marked with an
arrow, and the stars used to derive it are marked with ``M" in the last column.}
\label{t_RVs}
\begin{tabular}{l c c c r r l}
\hline
\hline
ID/NGC\,5999 & & $J$ & ($J-H$) & S/N & RV (km~s$^{-1}$) & Notes \\
\hline
\object{2MASS\,J15521389$-$5628139} & T & \ \ 8.68$\pm$0.01 & 0.77$\pm$0.03 & 40 & $-40\pm4$\ \ \ & M \\
\object{TYC\,8705-3241$-$1} & T & 10.71$\pm$0.02 & 0.10$\pm$0.03 & 30 & $-34\pm7$\ \ \ & M \\
\object{2MASS\,J15521176$-$5628045} & T & 11.08$\pm$0.03 & 0.11$\pm$0.05 & 30 & $-31\pm7$\ \ \ & M \\
\object{2MASS\,J15521358$-$5629086} & T & 11.12$\pm$0.02 & 0.12$\pm$0.04 & 25 & $-46\pm7$\ \ \ & M \\
\object{2MASS\,J15522168$-$5626451} & T & 11.29$\pm$0.03 & 0.16$\pm$0.04 & 20 & $-56\pm7$\ \ \ & $2\sigma$-clip \\
\object{2MASS\,J15522304$-$5628169} & A & 12.47$\pm$0.03 & 0.16$\pm$0.05 &  7 & $-26\pm25$ & \\
\object{2MASS\,J15520479$-$5628155} & A & 12.02$\pm$0.03 & 0.11$\pm$0.04 & 10 & $10\pm25$ & \\
\object{2MASS\,J15521710$-$5628042} & A & 13.24$\pm$0.02 & 0.21$\pm$0.04 &  3 & $-31\pm35$ & \\
\object{2MASS\,J15520566$-$5628020} & A & 12.41$\pm$0.02 & 0.12$\pm$0.03 & 10 & $-61\pm11$ & $2\sigma$-clip \\
\object{2MASS\,J15521445$-$5628073} & A & 10.43$\pm$0.03 & 0.63$\pm$0.04 & 20 & $0\pm5$\ \ \ & $2\sigma$-clip \\
\object{2MASS\,J15520421$-$5629071} & A & 11.56$\pm$0.03 & 0.15$\pm$0.04 & 15 & $-45\pm15$ & \\
\hline
All stars & & & & & $-33\pm7$\ \ \ & \\
$2\sigma$-clip & & & & & $-38\pm6$\ \ \ & \\
$2\sigma$-clip, only ``T" & & & & & $\Rightarrow -39\pm3$\ \ \ & \\
VBe\,3 & PN & & & & $-157\pm3$\ \ \ & \\
&&&&&&\\
&&&&&&\\
\hline
\hline
ID/NGC\,6067 & & & & & \\
\hline
Cluster &  & & & & $-40.0\pm0.2$ & \citep{Mermilliod08} \\
HeFa\,1 & PN & & & & $-141\pm3$\ \ \ \ & \\
&&&&&&\\
&&&&&&\\
\hline
\hline
ID/NGC\,2453 & & & & & & \\
\hline
\object{TYC\,6548-790-1} & T & \ \ 6.73$\pm$0.02 & 0.85$\pm$0.05 & 60 & $73\pm5$\ \ \ & binary? \\
\object{MSP\,159}        & T & 12.13$\pm$0.03 & 0.09$\pm$0.04 & 25 & $66\pm6$\ \ \ & pm-nm \\
\object{TYC\,6548-905-1} & T & 10.44$\pm$0.03 & 0.17$\pm$0.04 & 50 & $88\pm10$ & Emission, binary? \\
\object{TYC\,6548-507-1} & T & 11.59$\pm$0.02 & 0.04$\pm$0.03 & 30 & $-28\pm18$ & $2\sigma$-clip \\
\object{TYC\,6548-1643-1} & T & 11.71$\pm$0.02 & 0.09$\pm$0.03 & 20 & $70\pm9$\ \ \ & M \\
\object{MSP\,111}        & T & 11.81$\pm$0.02 & 0.10$\pm$0.03 & 20 & $66\pm8$\ \ \ & M \\
\object{MSP\,74}         & A & 11.87$\pm$0.03 & 0.21$\pm$0.04 & 20 & $78\pm6$\ \ \ & Emission \\
\object{MSP\,146}        & A & - & - & 15 & $112\pm13$ & $2\sigma$-clip \\
\object{MSP\,190}        & A & 13.30$\pm$0.05 & 0.12$\pm$0.09 & 10 & $44\pm34$ & pm-nm \\
\object{2MASS\,J07472558$-$2712019} & A & 12.81$\pm$0.02 & 0.13$\pm$0.04 & 10 & $101\pm27$ & \\
\hline
All stars & & & & & $72\pm12$ & \\
$2\sigma$-clip & & & & & $72\pm3$\ \ \ & \\
& & & & & $\Rightarrow 68\pm4$\ \ \ & \\
NGC\,2452 & PN & & & &  $65\pm1$\ \ \ & From literature (see text) \\
&&&&&&\\
&&&&&&\\
\hline
\hline
ID/NGC\,4463 & & & & & \\
\hline
\object{HD\,108719} & T & \ \ 7.84$\pm$0.03 & 0.06$\pm$0.04 & 110 & $-33\pm7$\ \ \ & \\
\object{CPD$-$64~1946} & T & \ \ 9.35$\pm$0.03 & 0.09$\pm$0.04 & 70 & $-16\pm5$\ \ \ & M \\
\object{2MASS\,12294384$-$6448130} & T & \ \ 9.73$\pm$0.02 & 0.05$\pm$0.03 & 50 & $-15\pm5$\ \ \ & M \\
\object{CPD$-$64 1944} & T & 10.49$\pm$0.02 & 0.06$\pm$0.03 & 35 & $-31\pm6$\ \ \ & \\
\object{2MASS\,12294705$-$6447464} & T & 10.51$\pm$0.02 & 0.04$\pm$0.03 & 40 & $-16\pm7$\ \ \ & M \\
\object{2MASS\,12301312$-$6447118} & A & 11.69$\pm$0.03 & 0.09$\pm$0.04 & 10 & $-39\pm10$ & \\
\object{2MASS\,12294673$-$6447092} & A & \ \ 9.97$\pm$0.02 & 0.65$\pm$0.03 & 15 & $-40\pm5$\ \ \ & \\
\object{2MASS\,12300426$-$6447427} & A & 12.44$\pm$0.04 & 0.05$\pm$0.06 & 6 & $-118\pm13$ & \\
\object{2MASS\,12295197$-$6447455} & A & 12.82$\pm$0.03 & 0.22$\pm$0.05 & 6 & $-66\pm18$ & \\
\object{CPD$-$64~1940} & & \ \ 9.73$\pm$0.02 & 0.05$\pm$0.03 & & $-35\pm4$\ \ \  & \ \ \ \ \ \ \citep{Hron85} \\
\object{CPD$-$64~1945} & & \ \ 8.82$\pm$0.02 & 0.10$\pm$0.04 & & $-16.\pm4$\ \ \  & M, \citep{Hron85} \\
\object{CPD$-$64~1943} & & \ \ 6.37$\pm$0.02 & 0.28$\pm$0.03 & & $-12.2\pm0.2$ & M, \citep{Mermilliod08} \\
\hline
& & & & & $\Rightarrow-15\pm2$\ \ \ & \\
He\,2-86 & PN & & & & $-11\pm$4\ \ \ \ & \\
\hline
\end{tabular}
\end{center}
\end{table*}


\section{Results}
\label{s_res}

\subsection{NGC\,5999 and VBe\,3}
\label{ss_5999}
Previous investigations of the cluster NGC\,5999 yielded consistent distances and reddenings ($d\approx$2~kpc,
$E(B-V)\approx$0.45), yet the cluster age estimates exhibit a $\sim50$\% spread \citep{Santos93,Piatti99,Netopil07}. No RV
measurements are available for the cluster stars. The PN VBe\,3 lies $\sim5\arcmin$ from the cluster center, which
places the PN within the cluster's coronal region \citep{Kholopov69}.  Star counts inferred from 2MASS photometry indicate
that the cluster's domination of the field terminates near $3.5 \arcmin$. A more comprehensive analysis of the radial
density profile is warranted \citep[e.g.,][]{Bonatto08}, but beyond the scope of this study because, as shown later in
this section, it is not required to assess the PN cluster membership. No estimates of $E(B-V)$, distance, or RV for the
PN were found in the literature.

Our observations targeted five candidate cluster members. Six additional objects fell in the slit and are included in our
analysis. The weighted mean of all eleven RVs is $V_R=-33\pm7$~km~s$^{-1}$, where the uncertainty is the
statistical error-on-the-mean.  However, a 2$\sigma$-clipping algorithm excludes three outliers. Those stars are tagged
with ``$2\sigma$-clip" in the last column of Table~\ref{t_RVs}. The weighted mean decreases to
$V_R=-38\pm6$~km~s$^{-1}$ after their exclusion, and the change introduced by the cleaning procedure is not
significant to within uncertainties. Among the selected targets, only 2MASS\,15522168$-$5626451 is excluded by this
procedure. The six additional stars exhibit larger uncertainties, and thus contribute less to the weighted mean. When only
considering the selected targets with the exclusion of 2MASS\,15522168$-$5626451, the weighted mean is $V_R=-39\pm3$~km~s$^{-1}$.
That result can be considered the best estimate of the cluster RV.

The RV of the PN is $V_R=-157\pm3$~km~s$^{-1}$. To our knowledge, there are no previous estimates available in the
literature, and ours is the first measurement of the RV of this nebula. The offset from the cluster value is sizable
($\approx 15\sigma$, defining the error on the difference as the quadratic sum of the uncertainties on the OC and PN
RVs). This proves strongly that the PN is not a cluster member. The association between VBe\,3 and NGC\,5999 is therefore a
chance alignment.

\subsection{NGC\,6067 and HeFa\,1}
\label{ss_6067}
NGC\,6067 is important for the distance scale since it hosts the classical Cepheid V340\,Nor \citep{Eggen83,Turner10}.
\citet{Turner10} cites a cluster distance of $d=1.7\pm0.1$~kpc, identical to the Cepheid distance established by
\citet[][]{Storm11} via the infrared surface brightness technique \citep{Fouque97}. The distance cited for PN HeFa\,1 is
highly uncertain, but \citet{Henize83} and \citet{Tylenda92} report a reddening of $E(B-V)=0.66$. That value is much higher
than the one established for V340\,Nor by \citet[][$E(B-V)=0.29\pm0.03$]{Turner10}. That discrepancy may indicate that the PN
is an unrelated background object, but this evidence alone is not conclusive because, as discussed in
Sect.~\ref{s_intro}, PNe can suffer from internal dust obscuration. The nebula is found at a large angular distance
from NGC\,6067 ($12\arcmin$), but the cluster is radially very extended. In fact, even the classical Cepheid QZ\,Nor at
$20\arcmin$ from the center is a cluster member \citep{Majaess13}, thus HeFa\,1 lies within the cluster boundaries.

\citet{Mermilliod87} established a precise RV for the cluster, namely $V_R=-39.3\pm1.6$~km~s$^{-1}$, updated to
$V_R=-39.99\pm0.18$~km~s$^{-1}$ by \citet{Mermilliod08}. Our result for the PN is $V_R=-141\pm3$~km~s$^{-1}$. As in
the previous case, this estimate is new in the literature, and it differs from the cluster value by $\approx 18\sigma$.
As a result, the PN is not a cluster member. A physical association between the PN and the OC is therefore excluded.

Incidentally, the results for this PN/OC chance alignment closely resemble those obtained for NGC\,5999 and VBe\,3
(Sect.~\ref{ss_5999}): the two OCs and PNe exhibit similar RVs ($V_R\approx -40$~km~s$^{-1}$ for the
clusters, $V_R\approx -150$~km~s$^{-1}$ for the PNe). Both pairs are projected toward the Galactic bulge
($l\approx 330\degr , b\approx -2\degr$), separated by only $3\fdg 5$. The PNe are likely members of the background bulge
population \citep[e.g.,][their Fig.~1]{Majaess07}.

\begin{figure}
\begin{center}
\includegraphics[width=9.5cm]{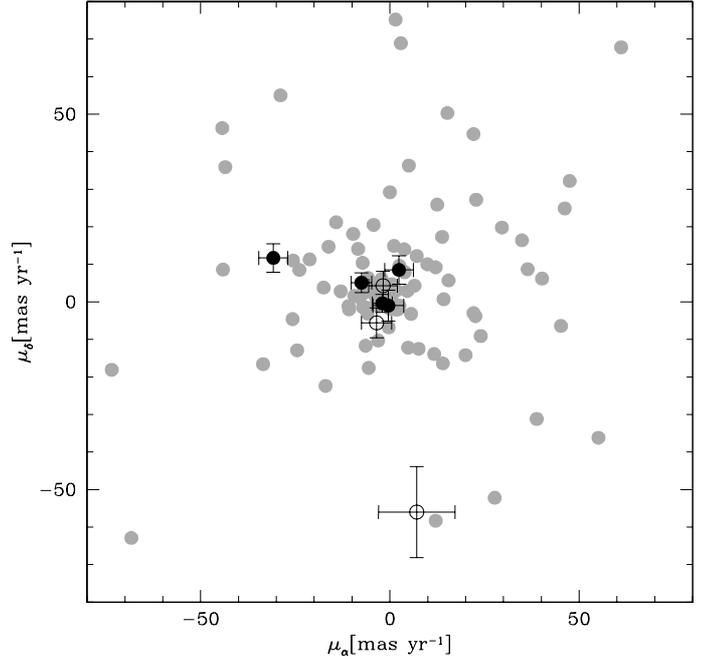}
\caption{Proper motions of stars within $1\farcm2$ from the center of NGC\,2453. The observed targets and additional
stars are indicated with full and empty dots, respectively.}
\label{f_pm2453}
\end{center}
\end{figure}

\subsection{NGC\,2453 and NGC\,2452}
\label{ss_2453}
Establishing a reliable distance for NGC\,2453 has proved problematic because the cluster is projected along the same line of
sight as the Puppis associations \citep{Majaess09a}. The resulting CMD is thus heavily contaminated by stars at different
distances and reddening. The cluster distance is known only within a factor of two, with some authors proposing
$d\approx$2.4--3.5~kpc \citep{Moffat74,Glushkova97,Dambis99,Hasan08}, and others finding a much higher value, $d\geq$5~kpc
\citep{Mallik95,Moitinho06}. As a consequence, the selection of our targets was not straightforward. There is agreement that
the cluster reddening is approximately $E(B-V)\sim$0.47--0.49, which is expected to slowly increase along this line of sight
beyond $d\approx$2~kpc \citep{Neckel80,Gathier86}.

NGC\,2452 is a massive PN \citep{Cazetta00}, whose progenitor must have been an intermediate-mass main-sequence star close
to the upper limit allowed for PN formation. This is consistent with the $\sim$40~Myr age of NGC\,2453 proposed by
\citet{Moffat74} and \citet{Moitinho06}, which implies a turnoff mass of $\approx$7~M$_\odot$. The nebula lies $\sim3.4$
cluster radii \citep[$\sim2\farcm5$,][]{Moffat74,Mallik95} from the center, i.e. well within the cluster corona. Depending on
the cluster distance, this corresponds to a physical separation of 6--12~pc. \citet{Zhang95} and \citet{Tylenda92}
established parameters for the PN of $d=3.0\pm 0.4$~kpc and $E(B-V)=0.36\pm 0.12$, respectively. \citet{Gathier86} proposed
$E(B-V)=0.43\pm 0.05$, roughly matching the cluster reddening, but they proposed a large difference between the cluster and
PN distances ($5\pm0.6$ and $\sim3.6\pm0.6$~kpc, respectively). However, the authors estimated these values from the
reddening-distance relation derived along the line of sight, but the reddening of the two objects agree within only
1$\sigma$. 

Our observations targeted six candidate members of NGC\,2453, and RVs were likewise
measured for four additional stars falling serendipitously in the slit. The weighted average of these ten measurements
yielded $V_R=72\pm12$~km~s$^{-1}$. After the exclusion of two outliers by means of a 2$\sigma$-clipping algorithm, as well as
of two other highly uncertain measurements (stars MSP\,190 and 2MASS\,J07472558$-$2712019), the result is
$V_R=72\pm3$~km~s$^{-1}$.  TYC\,6548-905-1 \citep[as first noted by][]{MacConnell81} and MSP\,74
\citep[first reported by][]{Moffat74} exhibit H$_\alpha$ emission, as indicated in the last column of Table~\ref{t_RVs}.
Their double-peaked asymmetric profile is not ideal for inferring RVs. In fact, these two objects have the highest RVs
among the six stars analyzed. 

Unfortunately, no weaker features suitable for RV measurements were detectable at the
resolution and S/N of our spectra in the restricted well-focused spectral range. Moreover, \citet{Moffat74} observed
double lines in the spectrum of TYC\,6548-905-1 and established $V_R=67\pm14$~km~s$^{-1}$ for the star, which is barely compatible
with our measurement despite the very large uncertainty. \citet{Mermilliod08} measured $V_R=85.24\pm 0.31$~km~s$^{-1}$ for
TYC\,6548-790-1, which differs from our estimate. They do not claim that the star is RV-variable, but their four
measurements are only consistent within 2.4$\sigma$. The resulting probability of RV variability is high
($\sim$93\%). Finally, the proper motion of MSP\,159 is offset from the other stars. The proper motions of program stars,
drawn from the PPMXL catalog \citep{Roeser10}, are given in Table~\ref{t_pm2453} and overplotted in
Fig.~\ref{f_pm2453} to that of all stars within $1\farcm 2$ of the cluster center. Unfortunately, the PPMXL catalog
provides no data for the stars MSP\,146 and TYC\,6548-905-1. The proper motions of stars near the cluster center are close
to zero, as expected for this distant object. All the observed stars agree with that distribution, except MSP\,159 and
MSP\,190, marked with ``pm-nm" in the last column of Table~\ref{t_RVs}. The latter object was also excluded from our
analysis because of its uncertain RV.

\begin{table}[t]
\begin{center}
\caption{Proper motions of program stars in NGC\,2453.}
\label{t_pm2453}
\begin{tabular}{l c c}
\hline
\hline
ID & $\mu_\alpha$ & $\mu_\delta$ \\
 & mas yr$^{-1}$ & mas yr$^{-1}$ \\
\hline 
TYC\,6548-790-1 & \ \ $-0.4\pm4.1$ & $-1.0\pm4.1$ \\
MSP\,74 & \ \ $-3.5\pm4.0$ & $-5.6\pm4.0$ \\
MSP\,159 & $-30.8\pm3.8$ & \ $11.7\pm3.8$ \\
MSP\,190 & \ \ \ \ \ \ \ $7.1\pm10.1$ & $-56.0\pm12.1$ \\
2MASS\,J07472558$-$2712019 & \ \ $-1.8\pm3.8$ & \ \ \ $4.3\pm3.8$ \\
TYC\,6548-507-1 & \ \ $-2.0\pm2.6$ & $-0.4\pm2.4$ \\
TYC\,6548-1643-1 & \ \ $-7.5\pm2.7$ & \ \ \ $5.1\pm2.6$ \\
MSP\,111 & \ \ \ \ \ $2.4\pm3.8$ & \ \ \ $8.5\pm3.8$ \\
\hline
\end{tabular}
\end{center}
\end{table}

The mean RV decreases after excluding some or all of these doubtful cluster members and can be as low as
$V_R=67\pm2$~km~s$^{-1}$ when the two emission-line objects and TYC\,6548-790-1 are excluded. A change of 5~km~s$^{-1}$
is not large, but this uncertainty must be taken into account. After excluding all the stars whose membership or
RV is uncertain for any reason, only two reliable measurements are left, whose mean velocity is $V_R=68\pm2$~km~s$^{-1}$.
Therefore, considering that different exclusion criteria lead to mean velocities in the range 67--72~km~s$^{-1}$, with
a statistical uncertainty of 2~km~s$^{-1}$, we conclude that the best estimate of the cluster RV is
$V_R=68\pm4$~km~s$^{-1}$.

Previous estimates for the RV of the PN NGC\,2452 have yielded $V_R=62.0\pm2.8$~km~s$^{-1}$ \citep{Meatheringham88},
$68.0\pm2.5$ \citep{Wilson53}, and $65\pm3$~km~s$^{-1}$ \citep{Durand98}. The weighted mean of these results is
reported in Table~\ref{t_RVs}. These values agree within 1--1.5$\sigma$ with
our estimate for the cluster. The RV analysis therefore supports cluster membership for the PN NGC\,2452 in NGC\,2453.
However, this result cannot be considered conclusive given the uncertainties discussed above. Additional research is
desirable. In particular, a precise assessment of the cluster distance in addition to a more extensive RV study would
settle the matter. We incidentally note that both NGC\,2453 and the PN NGC\,2452 are very young objects, and it is
reasonable to assume that they partake of the Galactic rotation. With a calculation similar to the one presented in
Sect.~\ref{ss_rvs}, we find that the RVs of both objects put them at about $d=6.5$~kpc from the Sun.

\subsection{NGC\,4463 and He\,2-86}
\label{ss_4463}
Our observations targeted five stars in NGC\,4463, and four additional objects fell into the slit. Two of our targets
were previously observed by \citet{Hron85}, who measured RVs for five cluster stars. They found
$V_R=-17.4\pm 3.0$~km~s$^{-1}$ for CPD$-$64~1946, in excellent agreement with our result, whereas their measurement for
HD\,108719 ($V_R=-10.0\pm 4.6$~km~s$^{-1}$) differs from the estimate established here. However, that peculiar object is
probably a binary system, as discussed in Sect.~\ref{ss_HD108719}. Another interesting star targeted by their
investigation, but not by ours, is the F-type supergiant CPD$-$64~1943. Their result ($V_R=-43.9\pm 5.8$~km~s$^{-1}$) is
very different from $V_R=-12.2\pm 0.2$~km~s$^{-1}$ found by \citet{Mermilliod08}. This last estimate is
particularly reliable, because it is the average of five measurements at different epochs.

\begin{figure}
\begin{center}
\includegraphics[width=9cm]{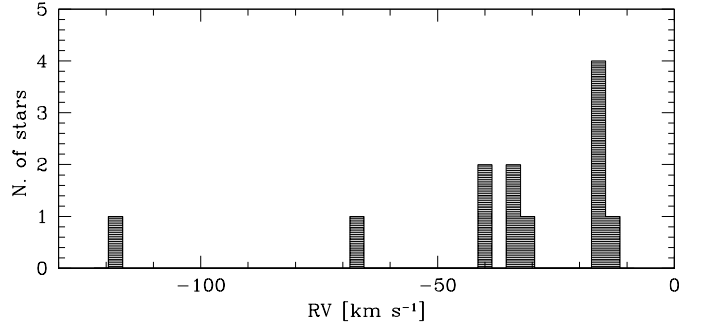}
\caption{Radial velocity distribution of the stars observed in NGC\,4463.}
\label{f_hist4463}
\end{center}
\end{figure}

\begin{figure}
\resizebox{\hsize}{!}{\includegraphics{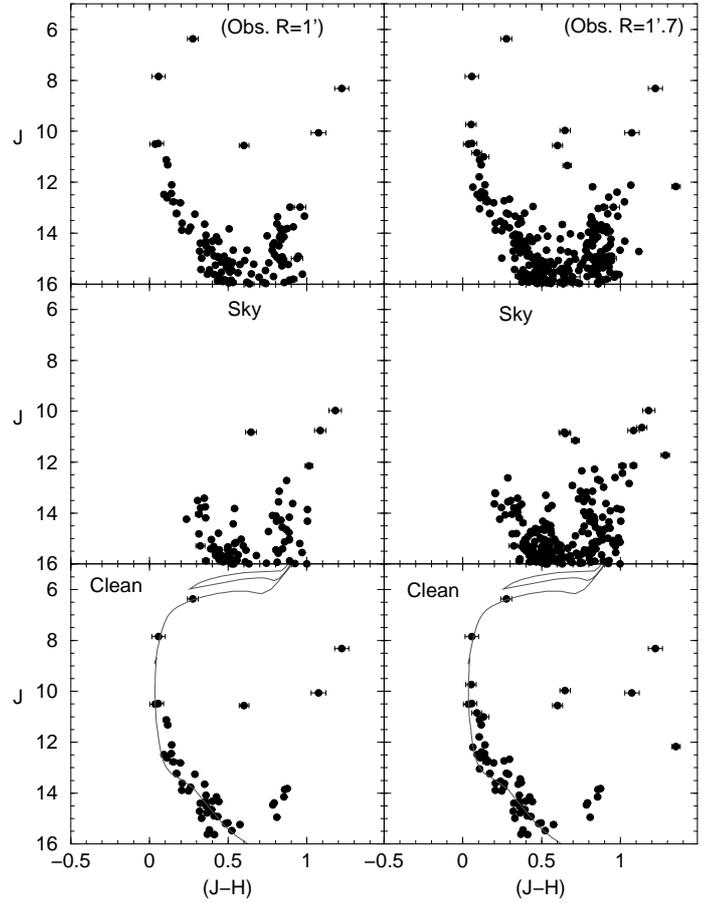}}
\caption{CMDs of NGC\,4463 extracted for $r=1\arcmin$ (left panels) and $r=1\arcmin.7$ (right). Panels show the observed
(top), same-area sky (middle) and decontaminated (bottom) CMDs. A 50\,Myr Padova isochrone of solar metallicity was
shifted by $E(J-H)=0.13$ and $d_\odot=1.54$\,kpc.}
\label{f_VVVCMD}
\end{figure}

The RV distribution of our stars is shown in Fig.~\ref{f_hist4463}, which includes the results of \citet{Hron85} for two
objects not in common with us, cited in Table~\ref{t_RVs}, and the measurement of \citet{Mermilliod08} for
CPD$-$64~1943. A group of five stars is found near $V_R\sim -15$~km~s$^{-1}$, comprising three of our selected targets,
and none of the additional stars. The other stars are scattered at more negative values, but tend to be distributed
between $-30$ and $-40$~km~s$^{-1}$. Half of the additional stars fall in this range. These objects have a much lower
cluster membership probability, because in the CMD they lie either off the cluster sequence or in the faint part of the
cluster MS ($J>12$) heavily contaminated by field stars (see the field CMD of Fig.~\ref{f_VVVCMD}). In fact, as discussed
in more detail in Sect.~\ref{ss_rvs}, \citet{Russeil98} found a kinematical group in the background of the cluster with mean
velocities between $-30$ and $-40$~km~s$^{-1}$. We conclude that the cluster RV is $V_R=-15.1\pm0.8$~km~s$^{-1}$. The
stars found to exhibit more negative RVs are considered field background stars.

The statistical uncertainty given here most probably underestimates the true error, because the RVs of four cluster stars
are identical within 1~km~s$^{-1}$, but the uncertainty on each of these measurements is $\sim$5~km~s$^{-1}$. Assuming this
typical measurement error as the standard deviation in the calculation of the error-on-the-mean is probably more representative
of the true uncertainty. In this way we fixed the final error as 2~km~s$^{-1}$.

The literature RV estimates for this cluster are summarized in Table~\ref{t_RV4463}. It must be noted that both
\citet{Kharchenko07} and \citet{Dias02} used the results of \citet{Hron85}, but \citet{Kharchenko07} obtained a lower value
because they averaged all the available measurements, while \citet{Dias02} excluded the two stars at
$V_R\approx-35$~km~s$^{-1}$. In fact, a similar distribution to the one observed in Fig.~\ref{f_hist4463} can be found among
the five stars of the \citet{Hron85} sample. \citet{Mermilliod08}, in contrast, based their results solely on CPD$-$64~1943.
The literature estimates for the RV of the PN He\,2-86 are also summarized in Table~\ref{t_RV4463}. Our result is
$V_R=-11\pm4$~km~s$^{-1}$, which agrees within 1$\sigma$ with all the previous estimates. The PN RV is similar to the cluster
velocity within 4~km~s$^{-1}$ ($0.9\sigma$), and is compatible with cluster membership within the errors.

\begin{table}[t]
\begin{center}
\caption{Literature RV estimates for NGC\,4463 and He\,2-86.}
\label{t_RV4463}
\begin{tabular}{l c c}
\hline
\hline
\multicolumn{3}{c}{NGC\,4463} \\
\hline
reference & N. of stars & RV (km~s$^{-1}$) \\
\hline
\citet{Kharchenko07} & 5 & $-24.52\pm6.37$ \\
\citet{Dias02} & 3 & $-14.6\pm4.0$ \\
\citet{Mermilliod08} & 1 & $-12.2\pm0.2$ \\
\hline
This work & 5 & $-15\pm2$ \\
\hline
\hline
\multicolumn{3}{c}{He\,2-86} \\
\hline
reference & & RV (km~s$^{-1}$) \\
\hline
\citet{Mendez81} & & $-7\pm7$ \\
 & & $-10\pm7 ^1$ \\
\citet{Acker92} & & $-7\pm2$ \\
\citet{Dopita97} & & $-8.9$ \\
\citet{Durand98} & & $-7.5\pm0.2$ \\
\citet{Garcia12} & & \ \ $-6\pm4 ^2$ \\
\hline
This work & & $-11\pm4$ \\
\hline
\end{tabular}
\end{center}
$^1$Central star RV. \\
$^2$Average of all single-line measurements with a 2$\sigma$-clipping algorithm selection.
The error indicates the line-to-line scatter.
\end{table}


\section{Fundamental parameters for NGC\,4463}
\label{s_vvv}

The RV of He\,2-86 is consistent with its membership in NGC\,4463. Near-IR photometric data collected by the
Vista Variables in the Via Lactea (VVV) ESO public survey may be employed to constrain the fundamental parameters of
NGC\,4463, to investigate the possible OC/PN connection in more detail.

The VVV survey is gathering near-IR data of the Galactic bulge and adjacent regions of the disk
\citep{Minniti10,Catelan11}. The central regions of the Milky Way are being surveyed four magnitudes deeper than 2MASS,
with excellent image quality and scale \citep[FWHM$<1\arcsec$, 0.34$\arcsec$/pix,][]{Saito10}. Sizable extinction may
shift a significant fraction of the main-sequence near/beyond the limiting magnitude of existing shallow surveys, thereby
complicating the efforts to apply isochrones and infer the cluster distance. Precise $JHK_s$ observations for stellar
clusters are consequently pertinent since obscuration by dust is less significant in the infrared. The VVV database is
therefore ideal for studying obscured stellar clusters in the Galactic disk \citep{Majaess12,Chene12}.

The age ($\tau$) and distance of NGC\,4463 are not well constrained in the literature. \citet{Kharchenko05} obtained
$d=1.05$~kpc and $\log{\tau}=7.97$, whereas the WEBDA web site\footnote{http://obswww.unige.ch/webda/navigation.html}
\citep{Mermilliod96,Mermilliod03} cites $\log{\tau}=7.505$ and \citet{Moffat73} finds $d=1.24$~kpc. Establishing the
parameters of this cluster using new and deep VVV $JHK_s$ photometry is thus desirable.

\subsection{VVV data and PSF-fitting photometry}
\label{ss_4463vvvdata}

NGC\,4463 is found in the VVV tile d005 \citep[see][for a description of VVV data, and the definition of VVV tiles
and pawprints]{Saito12}. We retrieved from the Vista Science Archive website (VSA)\footnote{http://horus.roe.ac.uk/vsa/}
the stacked pawprints (single 16--chip images) of this tile, collected in the five bands $YZJHK_s$, plus five additional
epochs in $K_s$. The data were acquired between February and April 2010 using the VIRCAM camera, which is mounted on the
VISTA 4m telescope at Paranal Observatory \citep{Emerson10}. The frames were pre-reduced (debiased, flat-fielded,
sky-subtracted, and stacked) at the Cambridge Astronomical Survey Unit (CASU)\footnote{http://casu.ast.cam.ac.uk/} with
the VIRCAM pipeline v1.0 \citep{Irwin04}. The weather conditions were optimal during the observations and fell within
the survey's constraints for seeing, airmass, and Moon distance \citep{Minniti10}.

Stellar PSF-fitting photometry was performed with the VVV-SkZ-pipeline \citep{Mauro12}, automated software based on
DAOPHOT \textsc{iv} \citep{Stetson87} and ALLFRAME \citep{Stetson94} that is optimized for extracting PSF photometry
from VVV images. We fed the pipeline all the single 2048x2048 pixel frames collected on chip \#4 of the VIRCAM array,
for a resulting $\sim20 \times 20\arcmin$ field. A total of sixty images were used, i.e. six frames in each of the five
bands and in each of the additional five epochs collected in the $K_s$ band. The $JHK_s$ instrumental magnitudes were
calibrated using 2MASS standards, following the procedure described by \citet{Chene12} and \citet{Moni11b}. The $YZ$
magnitudes were not calibrated owing to the lack of available standard stars in the field, and they will not be used in
our study.

\subsection{Field-star decontaminated VVV CMDs}
\label{ss_4463VVVdec}

Mitigating field-star contamination will facilitate an analysis of the CMDs. Field contamination is expected in low Galactic
latitude fields, such as the one hosting NGC\,4463. For this purpose we applied the statistical field-star
decontamination algorithm developed by \citet{Bonatto07}, adapted to exploit the VVV photometric depth in $J$, $H,$ and
$K_s$ \citep{Bonatto10,Borissova11,Chene12}. We combined several regions around NGC\,4463 to build the comparison field,
which maximizes the statistical representation of the field stars.

CMDs extracted within $r=1\arcmin$ and $r=1\farcm 7$ of cluster center are shown in the top panels of
Fig.~\ref{f_VVVCMD}, where a typical star cluster sequence detaches from the field contamination (middle panels).
Finally, the end products of the decontamination algorithm are the clean CMDs (bottom) in which most of the contaminant
stars have been removed. The clean CMDs are adequately fitted by a Padova isochrone \citep{Girardi00} of solar metallicity
and an age of 50~Myr (with an uncertainty of $\pm10$~Myr). Additional parameters for the isochrone setting are the apparent
distance modulus (m-M)$_J=11.3\pm0.1$, and the foreground reddening $E(J-H)=0.13\pm0.01$ that, with \citet{Cardelli89}
reddening relations, converts to $E(B-V)=0.42\pm0.03$ and $A_V=1.29\pm0.10$. Taken together, these parameters imply a
distance of $d=1.54\pm0.07$\,kpc for NGC\,4463.

\subsection{Reddening and distance}
\label{ss_4463redist}

\begin{figure}
\resizebox{\hsize}{!}{\includegraphics{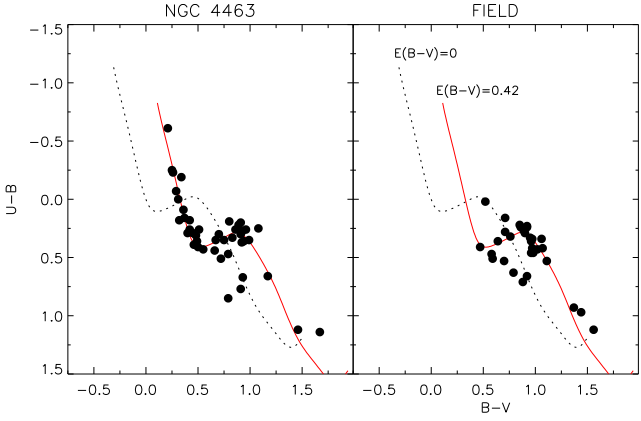}}
\caption{Optical color-color diagrams for stars lying within $r<1\farcm 3$ of the center of NGC\,4463 (left panel)
and in the surrounding field ($r>6\farcm 5$, right panel). Overplotted is the intrinsic relation of \citet{Turner89}
(dashed curve), and that relation shifted by $E(B-V)=0.42$ (solid curve).}
\label{f_colcol}
\end{figure}

Our first estimate of the cluster reddening and distance based on IR photometry was checked and refined by complementing
the VVV data with the optical $UBV$ data of \citet{Delgado07,Delgado11}. As shown in Fig.~\ref{f_colcol}, a $UBV$
color-color diagram for stars $r>6\farcm 5$ from the core is populated by objects typical of the field (i.e., later-type
objects), whereas objects within $r<1\farcm 3$ are dominated by cluster B-stars. The intrinsic $UBV$ relation of
\citet[][and references therein]{Turner89} was shifted along a reddening slope of
E$_{U-B}$/E$_{B-V}=0.72+0.02\times$E$_{B-V}$ \citep{Turner76}, to infer the reddening from stars lying within
$r<1\farcm 3$ of the cluster center. The resulting mean color excess is $E(B-V)=0.41\pm0.02$, which agrees
with our estimate based on VVV data.

The cluster distance may be established by means of multi-band CMDs, for the reddening previously determined and
shifting an intrinsic relation along the ordinate to match the target data. An empirical $JHK_s$ main-sequence calibration
established from deep 2MASS photometry and revised Hipparcos parallaxes for nearby stars is employed \citep{Majaess11b}.
That infrared calibration is comparatively insensitive to stellar age and metallicity, and it is anchored to seven
benchmark open clusters that exhibit matching $JHK_s$ and revised Hipparcos distances \citep{vanLeeuwen09}. The objective
is to avoid deriving distances using a single benchmark cluster (i.e., the Pleiades), thus potentially introducing a
possibly large systematic uncertainty. By employing seven benchmark clusters and shifting to the infrared where
metallicity effects are mitigated, that problem was avoided. For example, the Hipparcos parallax for the Pleiades
corresponds to a distance of $d=120.2\pm1.9$~pc \citep{vanLeeuwen09}, whereas HST observations imply $d=134.6\pm3.1$~pc
\citep[][see also \citealt{Majaess11b}, and discussion therein]{Soderblom05}.

\begin{figure}
\resizebox{\hsize}{!}{\includegraphics{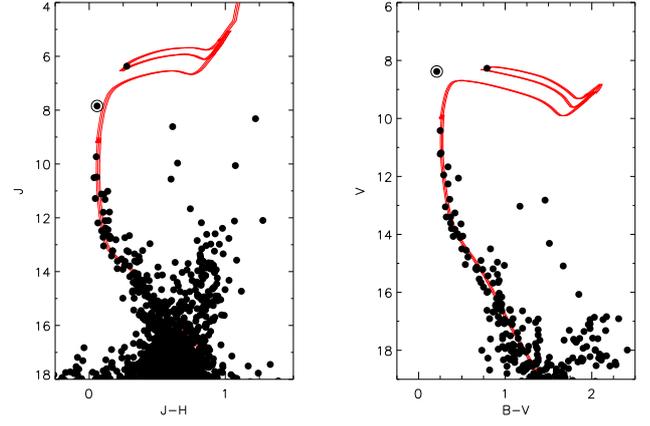}}
\caption{IR and optical CMDs for stars within $r<1\farcm 7$ from the center of NGC\,4463. The curves indicate the
isochrone described in the text. The empty circle highlights the position of the star HD\,108719.}
\label{f_dist}
\end{figure}

The $JH$ and $BV$ CMDs for stars $r<1\farcm 7$ from the cluster center are shown in Fig.~\ref{f_dist}. The VVV data were
supplemented by 2MASS photometry for the brightest stars saturated on the VVV images. The resulting distance to the cluster
is $d=1.55\pm0.10$~kpc, for a ratio of total to selective extinction $R_J=A_J/E(J-H)=2.75$ \citep{Majaess11c}.

\subsection{CPD$-$64~1943 and HD\,108719}
\label{ss_HD108719}

The cluster age proved more difficult to constrain than its distance, since any application of an isochrone is acutely
sensitive to the inclusion of the two brightest stars: CPD$-$64~1943 and HD\,108719. We therefore start by discussing them
in more detail.

CPD$-$64~1943 was classified by \citet{Fitzgerald79} as F5 Iab. \citet{Mermilliod08} measured a radial velocity
$V_R=-12.2\pm0.2$~km~s$^{-1}$ for this star, consistent with that established for the PN and the cluster mean.
\citet{Mermilliod08} do not indicate that the star is RV variable, but the measurement of \citet{Hron85} is noticeably
different ($-43.9\pm 5.8$~km~s$^{-1}$).

\begin{figure}
\resizebox{\hsize}{!}{\includegraphics{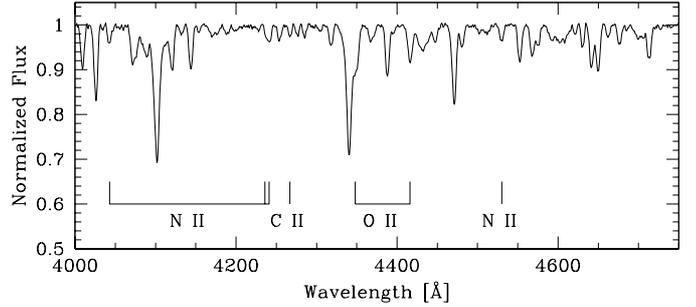}}
\caption{A portion of the spectrum of HD\,108719 collected on June 2012. The lines commented in the text are indicated.}
\label{f_specHD108719}
\end{figure}

HD\,108719 is the hottest star observed in the cluster area and, as such, its membership would strongly constrain the
cluster age. It is located at only $\sim 3\arcsec$ from the nominal cluster center. The initial mass function implies
that early-type stars are rare in a given field, and it is thus improbable from that perspective that the star is not a
cluster member. However, such an early-type star (spectral type B1, see below) must be massive
\citep[$\sim 11$~M$_{\sun}$, e.g.,][]{Ausseloos06,Harrington09}, and its expected lifetime is shorter than the time
required for the formation of the first PNe from the most massive progenitors of $\sim 8$~M$_{\sun}$
\citep[e.g.,][]{Weidemann00}. As a result, HD\,108719 cannot co-exist with He\,2-86, unless it is a blue straggler
\citep[BS, see][for a comprehensive analysis of BSs in open clusters]{Ahumada07}. Consequently, HD\,108719 does not
constrain the cluster age, and the star will be ignored in the
final isochrone fit. However, a BS candidate is a potentially interesting star, and we discuss its
cluster membership further. To better clarify its nature, we collected a blue spectrum of HD\,108719 on 2012 June 24 with
the same instrument but at lower resolution (R=2500), covering the blue range 3700--5265~\AA. A portion of this
spectrum is shown in Fig.~\ref{f_specHD108719}.

HD\,108719 exhibits radial velocity variations and is very likely a binary system: we measured
$V_R=-33\pm7$~km~s$^{-1}$ on April 13 and 10$\pm 11$~km~s$^{-1}$ on June~24, while \citet{Hron85} quoted
$-10.0\pm 4.6$~km~s$^{-1}$. Those results are consistent with a BS star, which has been affected by mass transfer.
Numerous multi-epoch observations are required to constrain the binary solution and to facilitate an evaluation of
membership based on radial velocities.

The star was classified as B1~III by \citet{Fitzgerald79}. An approximate estimate of the stellar parameters was obtained
by fitting the observed hydrogen and helium lines with a grid of synthetic
spectra obtained from model atmospheres of solar metallicity, computed with ATLAS9 \citep{Kurucz93}. The Balmer series from
H$_\beta$ to H$_{12}$, and the four $\ion{He}{I}$ lines at 4026~\AA, 4388~\AA, 4471~\AA, 4922~\AA, were included in the
fitting procedure. The routines developed by \citet{Bergeron92} and \citet{Saffer94}, as modified by \citet{Napiwotzki99},
were used to derive the stellar parameters by minimizing the $\chi^2$ of the fit \citep[for more details
on the fitting procedure, see][]{Moehler99,Moni12}. We thus obtained $T_\mathrm{eff}\approx$24\,000~K and
$\log{g}\approx$3.3, which are typical of a B1~II star. We also found that the star is a rapid rotator, with
$v \sin{i}\sim200$~km~s$^{-1}$. Unfortunately, this estimate must be considered as only approximate, because the assumed
solar metallicity could be inappropriate. Our low-resolution spectrum is not suitable for an accurate metallicity
measurement.

Following the spectroscopic classification criteria outlined by \citet{Gray09}, the spectrum for HD\,108719 is consistent
with a BN1~IIn star, because it exhibits enhanced nitrogen, whereas it is deficient in both oxygen and carbon. This
spectral type determination differs from the one cited by \citet{Fitzgerald79}. The star displays strong \ion{N}{ii}
features (3994, 4043, 4236-41, 4530~\AA) and weak \ion{O}{ii} (4348, 4416~\AA) and \ion{C}{ii} (4267~\AA) lines. The
luminosity class was inferred in part from the \ion{Si}{iii} triplet. The nitrogen enhancement could be associated with
rotational mixing. Alternatively, the spectral features may be suggestive of mass transfer from a companion or a merger,
which are consistent in part with the radial velocity variations.

\citet{Moffat73} cite $V=8.41$, $B-V=0.21$, and $U-B=-0.63$ for HD\,108719. According to \citet{Wegner07}, the absolute
magnitude of a B1~III, B1~II, and B1~Ib star is $M_V=-3.71$, $-$4.27, and $-$4.65, respectively. These values disagree
with the estimates of \citet{SchmidtKaler82}, namely $M_V=-4.4$, $-$5.4, and $-$5.8. For the B1~III spectral type
established by \citet{Fitzgerald79}, and the corresponding absolute magnitude from \citet{Wegner07}, the
distance to HD\,108719 is $d\sim1.5$ kpc. This coincides with the cluster distance established by us. However, the new
spectral type classification derived here implies $d\sim1.9$~kpc using the \citet{Wegner07} absolute magnitude or
$d\sim3.3$~kpc using the one established by \citet{SchmidtKaler82}. Given the large spread between these two values, we
derived an independent first-order determination of the absolute magnitude for a B1~II star. \citet{Bonanos09} compiled
a catalog of stars in the Large Magellanic Cloud (LMC) with optical photometry and spectroscopic classifications. Nine
stars with B1~II designations are in that sample and were analyzed to infer the absolute magnitude. The distance to the
LMC was adopted from \citet[][$\mu_0=18.43\pm0.03$, and see \citealt{Borissova09}]{Majaess11a}, who established it via
a universal Wesenheit template, which leverages the statistical weight of the entire variable star demographic to
establish precise ($<5$\%) distances. The template capitalizes upon HST \citep{Benedict07}, VLBA, and Hipparcos
geometric distances for SX~Phe, $\delta$~Scuti, RR~Lyrae, and for Type~II and classical Cepheid variables. A mean-color excess
of $E(B-V)=0.14$ is adopted for the B1~II stars and was inferred from Cepheid variables in the LMC \citep{Majaess09b}.
The resulting absolute magnitude for a B1~II is $M_V=-5.31\pm0.21\pm 0.63$, where the errors indicate the standard error
and standard deviation, respectively. That is consistent with the value reported by \citet{SchmidtKaler82}.
This estimate was based on LMC stars, which are on average more metal-poor than analogous Galactic stars by
$\Delta ([Fe/H])\approx -0.3$. However, the impact of metallicity is thought to be negligible in the absolute
magnitude of early-type stars, because most metal lines involved in line blanketing are absent. The small average
metallicity difference between LMC and Milky Way stars therefore cannot affect our estimate noticeably. This is confirmed by
inspecting Padova isochrones \citep{Girardi00} of metallicity Z=0.009 and Z=0.019, whose upper evolutionary sequences
($M_V<0$) differ by less than 0.1~mag. The resulting spectroscopic parallax for HD\,108719 (B1~II classification) is
$d\sim3.1$~kpc, which implies that the star is not a cluster member.

The reddening inferred for HD\,108719 using the intrinsic color cited by \citet{SchmidtKaler82} is $E(B-V) \sim0.47$.
This is only marginally more than what is derived for the cluster ($E(B-V)=0.41\pm0.02$, Sect.~\ref{ss_4463redist}),
which is approximately half as far. The latter result places firm constraints on the run of reddening along the line
of sight \citep{Neckel80}, namely that a foreground dust cloud is the primary source of obscuration for objects lying
between 1.5 and 3.2~kpc.

Finally, the possibility must be considered that HD\,108719 is not a hot supergiant, but a more exotic post-AGB
star. This scenario would explain its unusual location in the CMD, and its cluster membership status should be
reconsidered, although HD\,108719 should be ignored in the isochrone fitting even in this case. The lack of dust IR
emission from this star in the IRAS database \citep{iras88} argues against this hypothesis, but it does not necessarily
rule it out.

\subsection{Cluster age}
\label{ss_age}

A $\tau=45-50$~Myr Padova isochrone of solar metallicity  simultaneously matches the cluster sequence and the two
aforementioned bright stars best, as shown in Sect.~\ref{ss_4463VVVdec}. This would imply that the F-supergiant CPD$-$64~1943
is in the pre core-helium-burning phase and is traversing the HR-diagram for the first time. Stellar evolutionary models
predict that the first-crossing is rapid, and thus the probability of observing the star in that phase is less likely
than a second crossing (i.e., post onset of core-helium burning). However, as said in Sect.~\ref{ss_HD108719},
HD\,108719 should be ignored in the isochrone fitting. In this case, the best-fit isochrone is $\tau=65\pm10$~Myr,
which assumes that the F-supergiant initiated core-helium burning, and HD\,108719 is either a non-member or a BS. The
uncertainty stems from the offset of the isochrones that still fit the CMD reasonably well. The radial velocities,
however, do not unequivocally prove the cluster membership of CPD$-$64~1943.


\section{The cluster membership of He\,2-86}
\label{s_discuss}
As discussed in Sect.~\ref{s_intro}, radial velocity matches alone are a necessary but not sufficient condition for
assessing the physical connection of a PN/OC pair. Having derived the parameters of NGC\,4463, we can now gather the
available information to analyze the possible association of He\,2-86 with this cluster.

\subsection{Angular distance}
\label{ss_angdist}

\begin{figure}
\begin{center}
\includegraphics[angle=270,width=8.5cm]{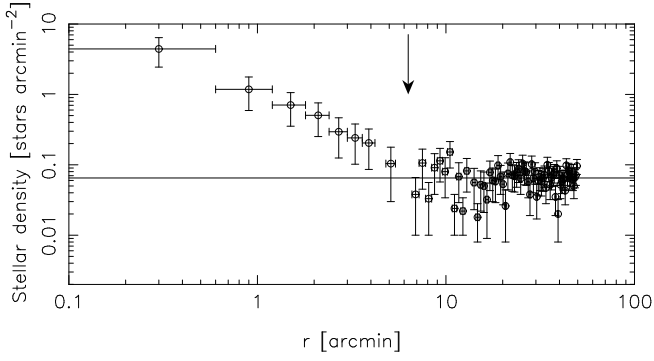}
\caption{Radial density profile of NGC\,4463, obtained from 2MASS data. The radial distance of He\,2-86 is indicated
with an arrow. The full line shows the field level, fixed as the average of all the points with $r>10\arcmin$.}
\label{f_radprof}
\end{center}
\end{figure}

The angular distance between He\,2-86 and the center of NGC\,4463 is $6\farcm4$, that at the cluster distance corresponds
to 2.9~pc. In Fig.~\ref{f_radprof} we show the cluster radial profile, constructed using 2MASS observations. The
cluster's brighter B-type stars, which rise prominently above the field in the color-magnitude diagram, are mostly
saturated in the deeper $VVV$ images, while available optical data are restricted to a smaller field of view. The
resulting profile indicates that the cluster's population appears significant to approximately r$\sim$8--10$\arcmin$
and that the PN is found well within the cluster boundaries. The angular distance of He\,2-86 to the cluster center is
therefore compatible with its membership to NGC\,4463.

\subsection{The progenitor mass}
\label{ss_progmass}

He\,2-86 is classified as an N-rich, \citet{Peimbert78} Type-I PN \citep{Pena13}. These PNe originate in high-mass
progenitors of 2--8~M$_{\sun}$, a fact that is consistent with the young age obtained for NGC\,4463 in Sect.~\ref{ss_age},
which implies a turnoff mass of $6.0\pm0.5$~M$_{\sun}$.

\begin{figure}
\begin{center}
\includegraphics[width=9.5cm]{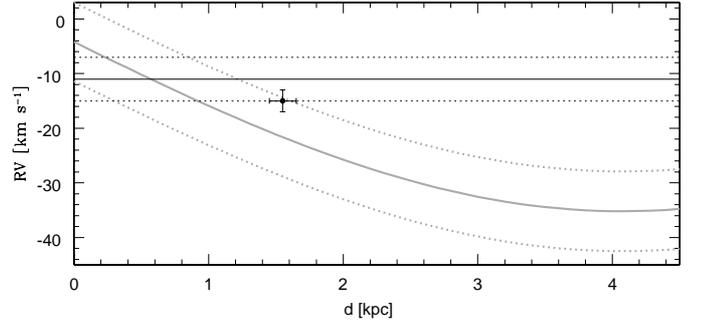}
\caption{Predicted radial velocity trend in the direction of NGC\,4463 from Galactic rotation, as a function of
distance from the Sun (gray solid curve). The gray dotted lines show the $1\sigma$ extremes when uncertainties in the
Galactic parameters are taken into account. The estimate derived for NGC\,4463 is indicated by the full dot with error
bars. The horizontal full and dotted lines show the RV of He\,2-86 and its 1$\sigma$ spread, respectively.}
\label{f_rvgal}
\end{center}
\end{figure}

\subsection{Radial velocities and Galactic rotation}
\label{ss_rvs}
As seen in Sect.~\ref{ss_4463}, our RV measurements for He\,2-86 and NGC\,4463 differ by only 4~km~s$^{-1}$ and are
consistent within 0.9$\sigma$. However, the errors are quite large, and a RV difference of $\approx$8~km~s$^{-1}$ is
still allowed within 1$\sigma$. More precise measurements are surely required to confirm the result.

In Fig.~\ref{f_rvgal}, we compare the RV of NGC\,4463 and He\,2-86 with the expected trend along their line of sight.
This was obtained assuming the rotation curve of \citet{Brand93}, the solar peculiar motion of \citet{Schonrich10},
$R_{\sun}=8\pm0.3$~kpc, and $V_{LSR}=220\pm20$~km~s$^{-1}$. The resulting figure confirms the early measurements of
\citet{Russeil98}, who found a group of stars with a mean velocity $-33$~km~s$^{-1}$ at $d=2$~kpc and a more distant
one that departs from pure circular motion, close to the tangential point ($d=4.4$~kpc), with $V_R=-40$~km~s$^{-1}$.

Both the OC and the PN are young, and it is reasonable to assume that they partake of the Galactic rotation to within a
few km~s$^{-1}$. In fact, the RV of NGC\,4463 matches the expectations for an object at $d$=1.55~kpc within the errors. On
the other hand, all the measurements (see Table~\ref{t_RV4463}) exclude that He\,2-86 is a background object. This trend
of the RV along the line of sight provides an upper limit of $d<$1.6~kpc, and the difference between the PN and the
Galactic disk velocity is $\Delta(RV)>$15~km~s$^{-1}$ (i.e., $>4\sigma$) if the object is assigned a distance $d>2$~kpc.

\subsection{Distance}
\label{ss_distance}

\begin{table}[t]
\begin{center}
\caption{Literature distance estimates for He\,2-86.}
\label{t_dPN}
\begin{tabular}{c l}
\hline
$d$ & reference \\
kpc & \\
\hline
3.54 & \citet{Cahn71} \\
2.6$\pm$1.8 & \citet{Maciel81} \\
1.4$\pm$0.4 & \citet{Amnuel84} \\
2.31 & \citet{Cahn92} \\
3.53$\pm$1.4 & \citet{vdSteene94} \\
3.58$\pm$1.5 & \citet{Zhang95} \\
1.4$\pm$0.9 & \citet{Tajitsu98} \\
5.1$\pm$3.1 & \citet{Cazetta00} \\
$>$2.7 & \citet{Cazetta01} \\
1.31$\pm$0.43 & \citet{Philips02} \\
8.6$\pm$2.6 & \citet{Philips04a} \\
3.85$\pm$0.18 & \citet{Philips04b} \\
\hline
\end{tabular}
\end{center}
\end{table}

The distance estimates for He\,2-86 in the literature, as summarized in Table~\ref{t_dPN}, span the very wide range from
1.3 to 5.1~kpc. It must be recalled, however, that they are all based on statistical methods whose results are not very
reliable for a single object. The resulting lower limit $d>$1.3~kpc excludes that the PN being in front of NGC\,4463.
By combining this information with the upper limit derived in Sect.~\ref{ss_rvs}, we can constrain the PN distance in the
narrow interval $d$=1.3--1.6~kpc. This matches remarkably well the cluster distance $d=1.55\pm0.10$~kpc derived in
Sect.~\ref{ss_4463redist}.

\subsection{PN size and age}
\label{ss_PNsize}
The apparent diameter of He\,2-86 is $\sim10\farcs7$g \citep{Westerlund67,Sahai11}.
For a distance of $d$=1.55~kpc, that corresponds to a physical radius of 0.04~pc, and for an expansion velocity of
22~km~s$^{-1}$ \citep{Acker92}, we obtain an expansion age of $\sim$1800~yr. A very similar result ($\sim$2200~yr) is
found by adopting the procedure of \citet{Gesicki06} and assuming their mass-averaged velocity of 14~km~s$^{-1}$ and a
mass-averaged radius of 0.032~pc (0.8 times the outer radius). If the PN is put at $d$=3~kpc (5~kpc), its radius becomes
0.08~pc (0.13~pc), with an expansion age of $\sim$3600~yr ($\sim$5800~yr). Again, an age greater by 25\% is obtained when
following the \citet{Gesicki06} method. All the values quoted for both the physical radius and the age are acceptable
for a \citet{Peimbert78} Type-I PN, hence these results do not put constraints on its cluster membership.

\subsection{Reddening}
\label{ss_redden}
\citet{Tylenda92} and \citet{Acker89} estimated $c$=1.86 for He\,2-86, which translates into $E(B-V)\approx$1.3~mag,
similar to the estimate of \citet{Frew08}. The deep discrepancy with the reddening of NGC\,4463 found in
Sect.~\ref{ss_4463redist} ($\Delta(E(B-V))\approx0.9$~mag) could indicate that the PN is a background object. However,
many arguments suggest that it is most likely because of heavy internal reddening.

First, the large difference between the expected interstellar reddening and the value inferred for the PN in not a
unique case in the literature. For example, similar objects can be found in \citet{Zagury05}, while \citet{Giammanco11}
find fourteen nebulae whose reddening exceeds the maximum reddening measured along their line of sight by more than one
magnitude. It is also worth remembering that \citet{Giammanco11} have shown that the PN reddening estimates cited above
are affected by an uncertainty of $\approx$0.3 magnitudes and by a possible systematic overestimate caused by
differential atmospheric dispersion. As the same authors note, the measurements obtained from radio data are free of
this effect, and they lead to systematically lower values. As an additional source of uncertainty, the extinction can
largely vary across dusty PNe \citep[e.g.,][]{Woodward92}, in particular for Type-I bipolar nebulae, such as the object
of this study \citep[e.g.,][]{Matsuura05}.

He\,2-86 is a multipolar, Type-I PN with a massive progenitor \citep{Sahai11,Pena13}, and significant internal
extinction may be expected for this kind of object from massive dusty envelopes \citep[e.g.,][]{Corradi95}.
\citet{Phillips98} showed evidence that internal extinction increases at smaller nebular radii, and it becomes
particularly relevant for small, young nebulae such as He\,2-86. According to his results, an internal reddening of
$E(B-V)=0.9$~mag is fully consistent with his simple model for the nebular radius $r=$0.04~pc, obtained from the
assumption that the PN is member of NGC\,4463 (Sect.~\ref{ss_PNsize}). Observational evidence for a relevant amount
of dust in He\,2-86 is also available. In fact, deep HST images reveal a ``dusty structure which
produces obscuration as it cuts across the lower lobes" \citep{Sahai11}. Both IRAS and WISE data
\citep{iras88,Wright10} consistently indicate that the PN is a bright IR source, and is indeed the brightest mid-IR
source in the local field in both the W3 (12$\mu$m) and W4 (22$\mu$m) bands. AKARI mid- and far-infrared data
\citep{Murakami07} corroborate the observations obtained from the aforementioned surveys. \citet{Tajitsu98} fit the
IRAS data with a modified blackbody function and derived a dust temperature of 134~K, and subsequently determined a
distance for the PN of 1.4~kpc. We note that the latter, although uncertain, is near the distance established here for
the cluster NGC\,4463. On the other hand, assuming a reddening discrepancy as evidence that He\,2-86 is behind
NGC\,4463 is inconsistent with other results presented here: the spectral classification of HD\,108719 and its
photometric properties indicate that the interstellar reddening beyond the cluster is nearly constant up to at least
$d=3.2$~kpc (Sect.\ref{ss_HD108719}), but a PN distance $d>3.2$~kpc is excluded by its RV (Sect.\ref{ss_rvs}).

In conclusion, all the available information points to high internal extinction for He\,2-86. As a consequence,
the reddening difference with NGC\,4463 is not informative about its possible cluster membership. On the other hand,
according to the study of \citet{Phillips98}, the PN/OC association hypothesis leads to a consistent picture, where
the observed extinction excess is very compatible with the inferred PN radius.

\subsection{The cluster membership}
\label{ss_memberesume}
As shown in this section, all the observational evidence points toward He\,2-86 being a young PN member of the cluster
NGC\,4463, with high internal reddening. In fact, while all the data agree with this scenario, the RV of the PN
excludes it being a background object, while all distance measurements in the literature exclude it from being the
foreground to the cluster. The observational evidence can only be reconciled by assuming the cluster distance for
the PN. He\,2-86 can be background to the cluster only if its orbit in the Galaxy differs noticeably from the disk
rotation. This scenario is very unlikely, because Type-I PNe such as He\,2-86 are young nebulae with a massive
progenitor, which formed between 20~Myr and 1~Gyr in the past, and their kinematics closely follow Galactic rotation
\citep{Pena13}. The high reddening of the PN cannot be interpreted as evidence of a background object, and its most
likely explanation is internal extinction that is high but not impossible in light of the available literature.


\section{Conclusions}
\label{s_conclusions}

We investigated the cluster membership of four PNe, which are found along the line of sight to known OCs. Our RV
measurements demonstrate that VBe\,3, and HeFa\,1 are two PNe located in the Galactic bulge, with no physical
relation with either NGC\,5999 or NGC\,6067, respectively. Conversely, the RV analysis supports cluster membership
for NGC\,2452 in NGC\,2453. However, this determination is not definitive, as NGC\,2453 is projected upon a
complex and dense stellar field. Our study of the cluster RV, similarly to the estimates of its reddening and
distance available in the literature, is likely to be affected by field star contamination.

Our analysis indicates that He\,2-86 is a young, internally highly reddened PN, probably a member of NGC\,4463.
The RV of the PN agrees within the uncertainties with the RV of the cluster, and this is the only consistent scenario that
can account for all the available observational evidence.

The distance and reddening for NGC\,4463 were estimated using new near-infrared photometry collected by the VVV survey, in
concert with optical observations. We find $d=1.55\pm0.10$~kpc and $E(B-V)=0.41\pm0.02$, and an age $\tau=65\pm10$~Myr.
The cluster distance and age can also be adopted for the PN He\,2-86. We find that NGC\,4463 hosts a PN and a
core-helium burning F-type supergiant, while the binary BS candidate HD\,108719 is probably a background
object, as inferred from its spectral classification, which places it beyond the cluster at $d\sim3.1$~kpc.


\begin{acknowledgements}
We thank the Cambridge Astronomical Survey Unit (CASU) for processing the VISTA raw data, and R. O. Gray and B. Skiff for
their comments and help. We thank the anonymous referee for the comments and suggestions. The authors acknowledge
support from the Chilean Centro de Astrof\'isica FONDAP No.~15010003, and the Chilean Centro de Excelencia en Astrof\'isica
y Tecnolog\'ias Afines (CATA) BASAL PFB/06. ANC acknowledge support from Gemini-Conicyt No.~32110005. Support for JB is
provided by Fondecyt Regular No.1120601. This investigation made use of data from the Two Micron All Sky Survey, which is a
joint project of the University of Massachusetts and the Infrared Processing and Analysis Center/California Institute of
Technology, funded by the National Aeronautics and Space Administration and the National Science Foundation. The authors
made extensive use of the SIMBAD and Vizier databases, operated at the CDS, Strasbourg, France.
\end{acknowledgements}


\bibliographystyle{aa}
\bibliography{PNe}

\end{document}